\documentclass[sort&compress
    ,final            % use final for the camera ready runs 
    ,numberedheadings % uncomment this option for numbered sections
    ,cmfonts          % add further options here if necessary
  ]
  {aipproc}

\usepackage{amsmath}
\usepackage{amssymb}
\usepackage{subfigure}
\newcommand{\be}{\begin{equation}}
\newcommand{\ee}{\end{equation}}
\newcommand{\bea}{\begin{eqnarray}}
\newcommand{\eea}{\end{eqnarray}}

\newcommand{\eqs}[1]{\begin{eqnarray}#1\end{eqnarray} }

\newcommand{\cf}[1]{{Figure~\ref{#1}}}
\newcommand{\cfs}[1]{{Figures~\ref{#1}}}

\newcommand{\eg}{{\it e.g.}}
\newcommand{\nn}{\nonumber}
\renewcommand{\cal}{\mathcal}

\layoutstyle{6x9}

\begin{document}

\title{$J/\psi$ production at the Tevatron and RHIC from $s$-channel cut}

\author{J.P. Lansberg}{
  address={Institut f\"ur Theoretische Physik, Universit\"at Heidelberg,\\ Philosophenweg 19, 69120 Heidelberg, Germany}}
\author{H. Haberzettl}{
address={Center for Nuclear Studies,
 Department of Physics, The George
Washington University, Washington, DC 20052, USA\\
E-mail: lansberg@tphys.uni-heidelberg.de,helmut@gwu.edu}}

\begin{abstract}

We report on our recent evaluation of the $s$-channel cut contribution to $J/\psi$ hadro-production.
We show that it is likely significantly larger than the usual cut contribution of the colour-singlet
model (CSM), which is known to underestimate the experimental measurements. Here the
$s$-channel cut develops for configurations with off-shell quarks in the bound state. A correct 
treatment of its contribution requires the introduction of a four-point function, partially
constrained by gauge invariance and limiting behaviours at small and large momenta.
When the unconstrained degrees of freedom are fixed to reproduce the Tevatron data, we show
that RHIC data are remarkably well reproduced down to very low transverse momenta $P_T$ without
need of resummation of initial-state gluon effects. This unique feature might be typical of $s$-channel
cut contribution. 

\end{abstract}

\maketitle

%\renewcommand{\thefootnote}{\fnsymbol{footnote}}
%\footnotetext{}
%\renewcommand{\thefootnote}{\arabic{footnote}}

\section{Introduction}

More than ten years after the first measurements by the CDF Collaboration of the {\it direct}
production of $J/\psi$ and $\psi'$ at $\sqrt{s}=1.8$
TeV~\cite{Abe:1997jz,Abe:1997yz} we are still facing disagreements between
theoretical predictions from the various available models and experimental studies of the
cross section and the polarisation from the Tevatron and RHIC 
(for reviews see~\cite{review}).

CDF~\cite{Abulencia:2007us} recently confirmed their previous
polarisation measurement~\cite{Affolder:2000nn} showing an unpolarised or
slightly longitudinally polarised prompt $J/\psi$ yield. This has reinforced the doubts cast on
the dominance of the Colour Octet Mechanism (COM) coming from the application
of NRQCD~\cite{Bodwin:1994jh}. 
At the same time, many new results became available, 
\eg~the long-awaited NLO QCD corrections to the CSM~\cite{Campbell:2007ws} 
--showing significant enhancement of the cross section, an
up-to-date proof~\cite{Nayak1} of NRQCD factorisation; an improved treatment
of NRQCD factorisation in fragmentation regions where three heavy quarks have similar
momenta~\cite{Nayak:2007mb} and last but not least a recent evaluation of the dominant $\alpha_S^5$ (NNLO)~\cite{Artoisenet:2008fc}
correction to $\Upsilon(nS)$ production, the latter solving the longstanding conflict between the 
experimental measurements from the Tevatron at {\it mid and large} $P_T$~\cite{Acosta:2001gv,Abazov:2005yc} and the prediction from the CSM~\cite{CSM_hadron}.

Considering that none of the existing theoretical approaches could 
reproduce all available experimental data,  we undertook in~\cite{Lansberg:2005pc} a
systematic study of the cut contributions due to off-shell and non-static
quarks. In particular, we questioned the assumption of the CSM that takes the
heavy quarks forming the quarkonium ($\mathcal{Q}$) as being
on-shell~\cite{CSM_hadron}. If they are not, the usual $s$-channel cut
contributes to the imaginary part of the amplitude and need to be considered on
the same footing as the CSM cut.

Current conservation for such off-shell configuration responsible for the $s$-channel cut
 imposes the introduction
of an additional four-point function, or contact
current~\cite{Drell:1971vx},  accounting for the
interactions between the $c\bar{c}$ pair emitting the external gluon. In fact,
this mechanism arises because of the possibility that the outgoing gluon 
is emitted by the particle interacting in the dressed $c-\bar{c}-J/\psi$ 
vertex (see Figure~\ref{fig:illus_GI_break} (a)), as depicted in Figure~\ref{fig:illus_GI_break} (b). The pair of on-shell 
quarks that makes the final $J/\psi$-gluon state is now in a colour-octet state which thus recovers the
necessity for such configurations as a natural consequence of restoring gauge
invariance.

Although current conservation imposes the introduction of such a 4-point vertex, it does
not enable to relate it univocally to the 3-point one.
Yet, there exist certain  minimal requirements~\cite{Kazes:1959,Drell:1971vx}
which such a 4-point function should satisfy. The 4-point function
proposed in~\cite{Lansberg:2005pc} provided a conserved current but was not
entirely satisfactory since it contained  poles (by construction similar to the basic direct
and crossed contributions), and such poles for the contact current are
unphysical and therefore should be avoided~\cite{Drell:1971vx}.

Another caveat to avoid was formerly identified by Drell and Lee~\cite{Drell:1971vx}.
Indeed, the minimal substitution prescription $\partial^\mu \to \partial^\mu +i Q A^\mu$  
($Q$: charge, $A^\mu$:
vector potential) in an effective Lagrangian corresponding to the dressed
hadronic vertex is
deficient in that it violates the high-energy scaling behaviour, because in
avoiding poles for the 4-point function, it partially replaces the true
momentum dependence of the vertices by constants. 

Such an issue can be easily avoided in the approach which we shall follow and which was
applied to pion photoproduction
processes~\cite{Haberzettl:1997jg,Haberzettl:1998eq,Davidson:2001rk,Haberzettl:2006fsi}.
As we shall show in the following, it is hence possible to build a 4-point vertex
 encompassing two limiting behaviours, when
the final-state gluon is soft or hard~\cite{Haberzettl:2007kj}. In turn, we shall show 
that this enables to reproduce experimental data from the Tevatron up to mid $P_T$ 
by adjusting the unconstrained parameters of the 4-point vertex  and hence to
get a remarkable agreement with data from RHIC down to low $P_T$.

\section{Our approach}

\subsection{The three-point function}

\begin{figure}[h!]\centering
\includegraphics[width=.6\columnwidth,clip=]{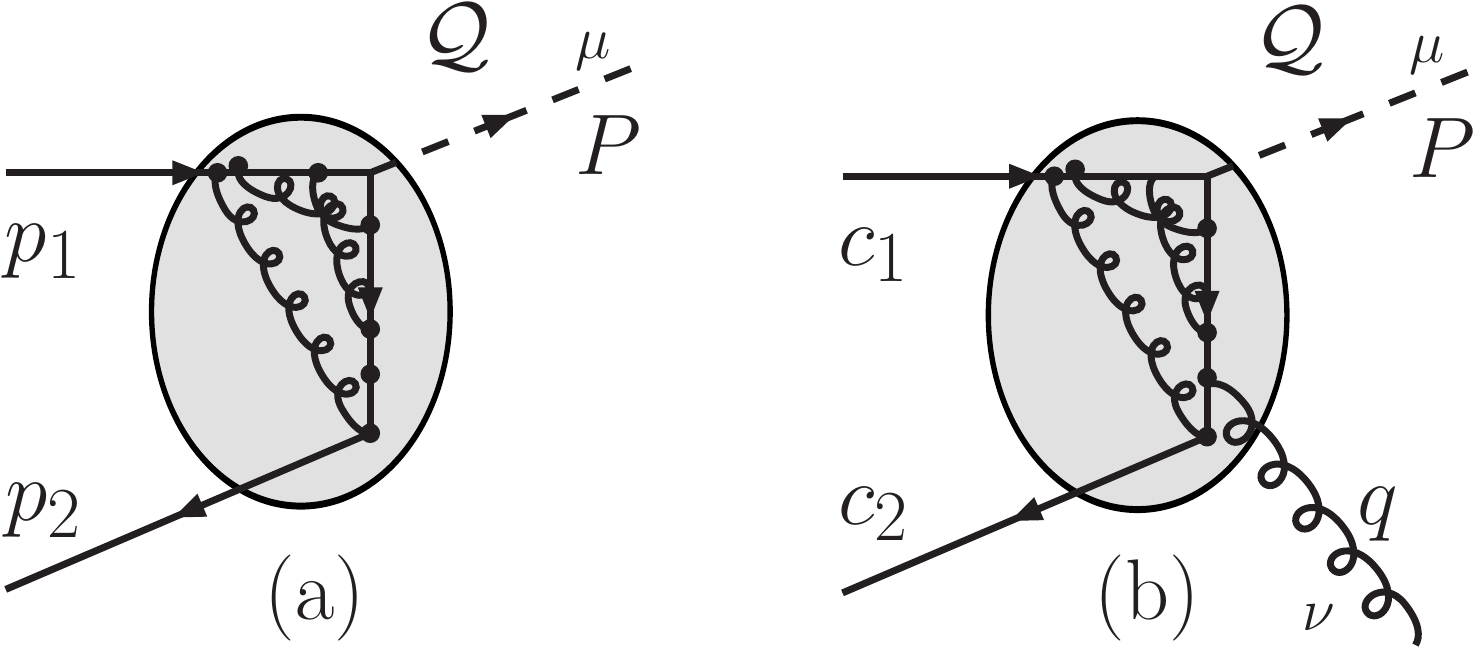}
 \caption{\label{fig:illus_GI_break}%
Illustration of the mechanisms (a) contributing to the dressing of the 3-point
function $\Gamma^{(3)}$ and (b) responsible for the 4-point function
$\Gamma^{(4)}$: the external gluon is attached here within gluon loops of the
dressed vertex, thus producing a %one-particle irreducible
diagram without poles
and with a kinematic behaviour genuinely different from the initial 3-point
vertex.}
\end{figure}

We shall follow the approach developed
in~\cite{Lansberg:2005pc}, where the transition $q\bar q\rightarrow {\mathcal
Q}$ is described by the 3-point function
\begin{equation}\label{vf}
\Gamma^{(3)}_{\mu}(p,P) = \Gamma(p,P) \gamma_\mu~,
\end{equation}
where  $P\equiv p_{1}-p_{2}$  and $p\equiv(p_{1}+p_{2})/2$ are the total and
relative momenta, respectively, of the two quarks bound as a quarkonium state,
with $p_1$ and $p_2$ being their individual four-momenta. Ansatz (\ref{vf})
amounts to representing the vector meson as a massive photon with a non-local
coupling. The generic picture of the physical origin of the dressed vertex function
$\Gamma(p,P)$ is given in Figure~\ref{fig:illus_GI_break}(a). 
In the present work, we describe the relative-momentum distribution
$\Gamma(p,P)$ of the quarks phenomenologically as a Gaussian form, function
of the square of the
relative c.m. 3-momentum $\vec p$ of the quarks, which can be written in a 
Lorentz invariant form as $\vec p^{\,\, 2}=
-p^2+\frac{(p.P)^2}{M^2}$. Explicitly, we have
\eqs{ \Gamma(p,P)=N e^{-\frac{\vec p^{\,2}}{\Lambda^2}} 
,}
with a normalisation $N$ --fixed by the 
leptonic-decay width~\cite{Lansberg:2005pc}--  and a size parameter $\Lambda$, which
can be obtained from studies in relativistic quark models~\cite{Lambda}.

\subsection{The four-point function: minimal substitution}

Instead of directly employing the generalised Ward--Takahashi 
relations~\cite{Kazes:1959} for the complete current, we will rather make use
of an equivalent condition for the contact current similarly to what was done 
in~\cite{Haberzettl:1997jg} in the pion-photoproduction case.

\begin{figure}[t!]\centering
  \includegraphics[width=.65\columnwidth,clip=]{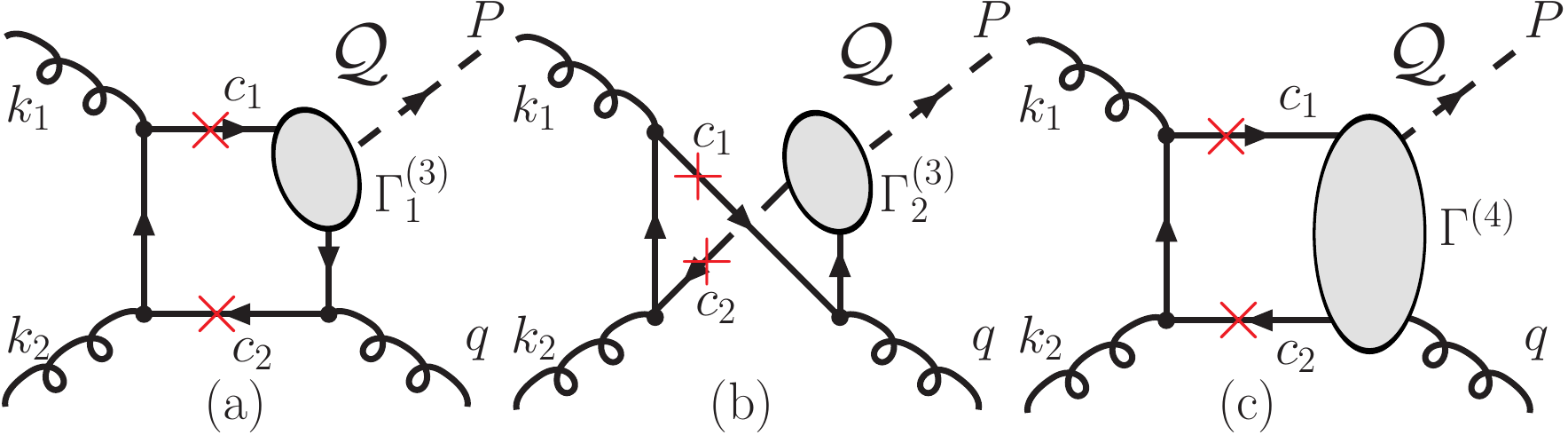}
  \caption{\label{fig:Jpsibox}%
  (a) \& (b) Leading-order (LO) $s$-channel cut diagrams contributing to $gg\to
J/\psi g$ with direct and crossed box diagrams employing  the
$c-\bar{c}-J/\psi$ vertex.  The crosses indicate that the quarks are on-shell.
(c) Box diagram with the 4-point vertex $c-\bar{c}-J/\psi-g$. 
Diagrams with reversed quark lines are not shown.}
\end{figure}

First let us write the 4-point function depicted in
Figure~\ref{fig:illus_GI_break}(b) as
\begin{equation}
\Gamma^{(4)} =-ig_s T^{a}_{ik} M_c^\nu \gamma^\mu~,
\label{eq:4pointfct}
\end{equation}
 where $g_s$ is the strong coupling constant, $T^{a}_{ik}$ the
colour matrix, and $\mu$ and $\nu$ are the Lorentz indices of the outgoing
$J/\psi$ and gluon, respectively. For simplicity, we have suppressed all
indices on the left-hand side. The $c-\bar{c}-J/\psi$ vertex function
$\Gamma^{(3)}$ with the kinematics of the direct graph is denoted here by
$\Gamma_1$ and for the crossed graph by $\Gamma_2$, {\it i.e.}, $\Gamma_1 =
\Gamma\left(c_1-\frac{P}{2},P\right)$ and $\Gamma_2 =
\Gamma\left(c_2+\frac{P}{2},P\right)$, as shown in Figures.~\ref{fig:Jpsibox}(a)
and (b). The gauge-invariance condition for the contact current $M_c^\nu$ for
the outgoing gluon with momentum $q$ reads now
\begin{equation}
q_\nu M_c^{\nu} = \Gamma_1-\Gamma_2
 \label{gipcond}
\end{equation}

The contact current can be now constructed as 
usual~\cite{Haberzettl:1997jg,Haberzettl:1998eq,Davidson:2001rk,Haberzettl:2006fsi}
in terms of an auxiliary function $F=F(c_1,c_2,q)$ which contains the remaining unconstrained
degrees of freedom of the problem. This gives
\begin{equation}
M_c^\nu = \frac{(2c_2+q)^\nu\left(\Gamma_1-F\right)}{(c_2+q)^2-m^2}
 +\frac{(2c_1-q)^\nu\left(\Gamma_2-F\right)}{(c_1-q)^2-m^2}~,
 \label{Mc}
\end{equation}
where we take  $c_1^2=c_2^2=m^2$ and $P^2=M^2$ from the beginning, with $m$ and
$M$ being the masses of the quark and the  $J/\psi$, respectively. One easily
verifies that this additional contact current satisfies the gauge-invariance
condition (\ref{gipcond}).

$F(c_1,c_2,q)$ has now to be chosen so that the current (\ref{Mc})
satisfies crossing symmetry (\textit{i.e.}, symmetry under the exchange
$c_1\leftrightarrow -c_2$) and is free of singularities. Defining the 
constant $\Gamma_0$ as  the (unphysical) value of the momentum distribution $\Gamma(p,P)$
when all three legs of the vertex are on their respective mass shells, we should 
then have $F(c_1,c_2,q)=\Gamma_0$ when $(c_2+q)^2=m^2$
or $(c_1-q)^2=m^2$.

A priori, $F=\Gamma_0$ everywhere should be satisfactory. However, this corresponds to
the minimal substitution discussed by Drell and Lee~\cite{Drell:1971vx} and later by 
Ohta~\cite{Ohta:1989ji}. As we mentioned above, this
does not provide the correct scaling properties at large energies, which means
within the present context that $F=\Gamma_0$ would not lead to the expected
$P_T$ scaling of the amplitude. Numerically, this choice overshoots the
experimental data by more than one order of magnitude at $P_T=20$ GeV as shown 
on~\cf{fig:dsdpt-gamma_0}.

\begin{figure}[h!]\centering
  \includegraphics[width=.75\columnwidth,clip=]{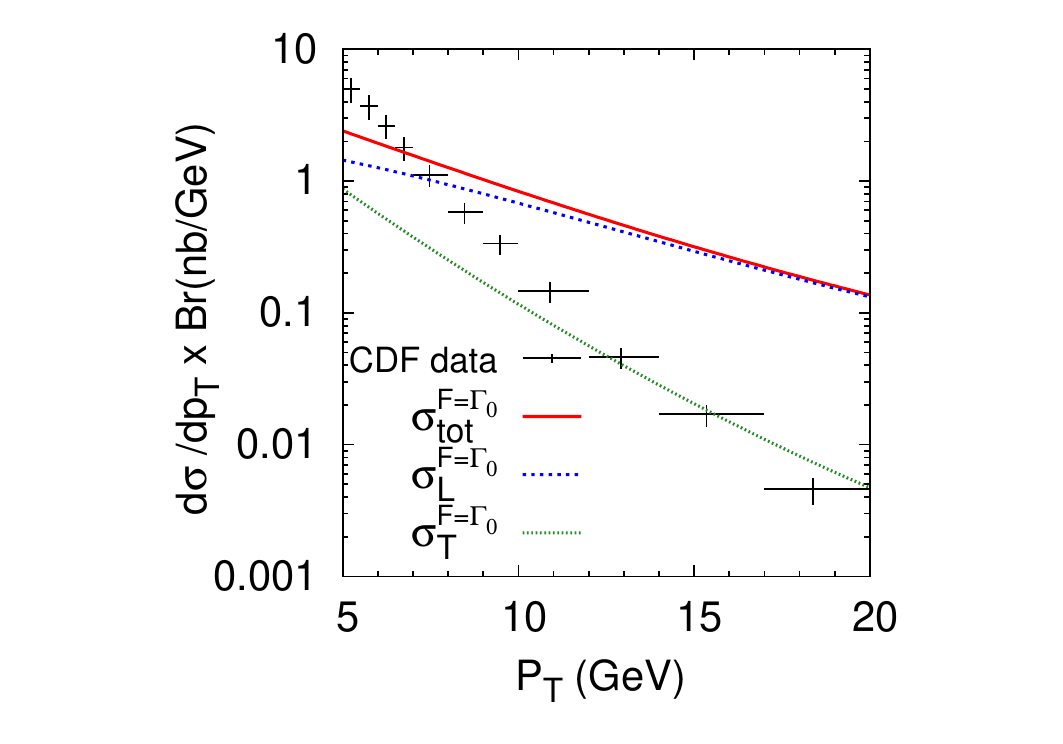}
  \caption{\label{fig:dsdpt-gamma_0}%
Comparison between the $J/\psi$ production cross section from $s$-channel cut
obtained with the minimal substitution ($F=\Gamma_0$) for the 4-point function 
and the CDF data~\protect\cite{Abe:1997yz}. See below for details on the 
cross-section evaluation.}
\end{figure}

\subsection{The four-point function: our proposal}

To avoid such abnormal scaling behaviour, we have to impose, 
in the large relative-momentum region, that the contact term and
therefore the function $F(c_1,c_2,q)$ exhibit a fall-off similar to the
3-point vertex functions.

The simplest crossing-symmetric choice satisfying this property
is~\cite{Davidson:2001rk}
\begin{equation}
   F= \Gamma_1+\Gamma_2 -\frac{\Gamma_1\Gamma_2}{\Gamma_0}~.
   \label{eq:scaling}
\end{equation}
The solution we propose here is to build $F(c_1,c_2,q)$ from these two limiting
cases:
\begin{align}
   F=& \Gamma_0 & \hbox{ at low momentum,} \nn\\
   F=& \Gamma_1+\Gamma_2 -\frac{\Gamma_1\Gamma_2}{\Gamma_0} & \hbox{ at large momentum}.\nn
   \label{eq:scaling2}
\end{align}

 To this end, it is practical to choose the following simple Ansatz
\begin{equation}
F(c_1,c_2,q)= \Gamma_0-h(c_1\cdot
c_2)\frac{\left(\Gamma_0-\Gamma_1\right)\left(\Gamma_0-\Gamma_2\right)}{\Gamma_0}~,
\end{equation}
where the (crossing-symmetric) function $h(c_1\cdot c_2)$  rises to become
unity for large relative momentum. 

Note that $F= \Gamma_0$ at the poles is satisfied independently of $h(c_1\cdot c_2)$. Indeed,
it is multiplied on the right by a factor which vanishes  at the poles since either $\Gamma_1=\Gamma_0$ or
$\Gamma_2=\Gamma_0$.  A phenomenological choice for the
inter\-polating function $h(c_1\cdot c_2)$ we can then propose is
\begin{equation}
h(c_1\cdot c_2)= 1- a\frac{\kappa^2}{\kappa^2-(c_1\cdot c_2+m^2)}~,
 \label{eq:interpolate}
\end{equation}
with two parameters, $a$ and $\kappa$. This choice is in no way unique: 
in a manner of speaking, this choice is
simply a way of parameterising our ignorance by employing minimal properties of
$\Gamma^{(4)}$. 

Other choices could be analysed but our conclusions, that $s$-channel cut 
contribution can be large and can indeed help to reproduce the experimental 
data, would not be affected.

\section{Results}

For the Tevatron and RHIC kinematics, the direct $J/\psi$ are
produced by gluon fusion and a final-state gluon emission is required to
conserve $C$-parity and provide the $J/\psi$ with its $P_T$. The relevant
diagrams for the  $s$-channel cut of the LO gluon fusion process are shown on~\cf{fig:Jpsibox}. 
We use the same normalisation of
$\Gamma^{(3)}$ as in~\cite{Lansberg:2005pc}, $m_c=1.87$ GeV and $\Lambda=1.8$ GeV.
As shown in~\cite{Lansberg:2005pc}, $\Lambda$ can be varied between 1.2 and 2.2 GeV 
without affecting much the results. The same statement holds here.

The partonic differential cross section obtained from the amplitude calculated
from our model (see~\cite{Lansberg:2005pc} for details) is summed over the
gluon polarisations, \textit{i.e.},
\begin{equation}
\frac{d \sigma_{r}}{d\hat t}=\frac{1}{16\pi \hat s^2}
 \sum_{p,q,s=T_1,T_2}|\overline{{\cal M}^{pqrs}}|^2
      ~,\quad r=L,T_1,T_2~,
\end{equation}
where $|\overline{{\cal M}^{pqrs}}|^2$ is the squared polarised partonic
amplitude for $g_p(k_1) g_q(k_2) \to {\cal Q}_r(P) g_s(q)$ averaged only over
colour for polarised cross sections. Here,  $p$, $q$, $r$ and $s$ are the
helicities of the respective particles, and $\hat s=(k_1+k_2)^2$, $\hat
t=(k_2-q)^2$ and $\hat u=(k_1-q)^2$ are the Mandelstam variables for the
partonic process. The relation to the double-differential polarised cross
section in transverse momentum $P_T$ and rapidity $y$ then is given by
\begin{equation}
\frac{d\sigma_r}{dy\,dP_T}=\int_{x_1^{{\rm min}}}^1 dx_1 \frac{2 \hat s P_T
g(x_1) g\left(x_2(x_1)\right)} {\sqrt{s}(\sqrt{s}x_1-E_T
e^{y})}\frac{d\sigma_r}{d\hat t}~.
\end{equation}
In the present calculations, we use the LO gluon distribution $g(x)$
of~\cite{Pumplin:2002vw}.

\begin{figure}[t!]\centering
\includegraphics[width=.6\textwidth,clip=]{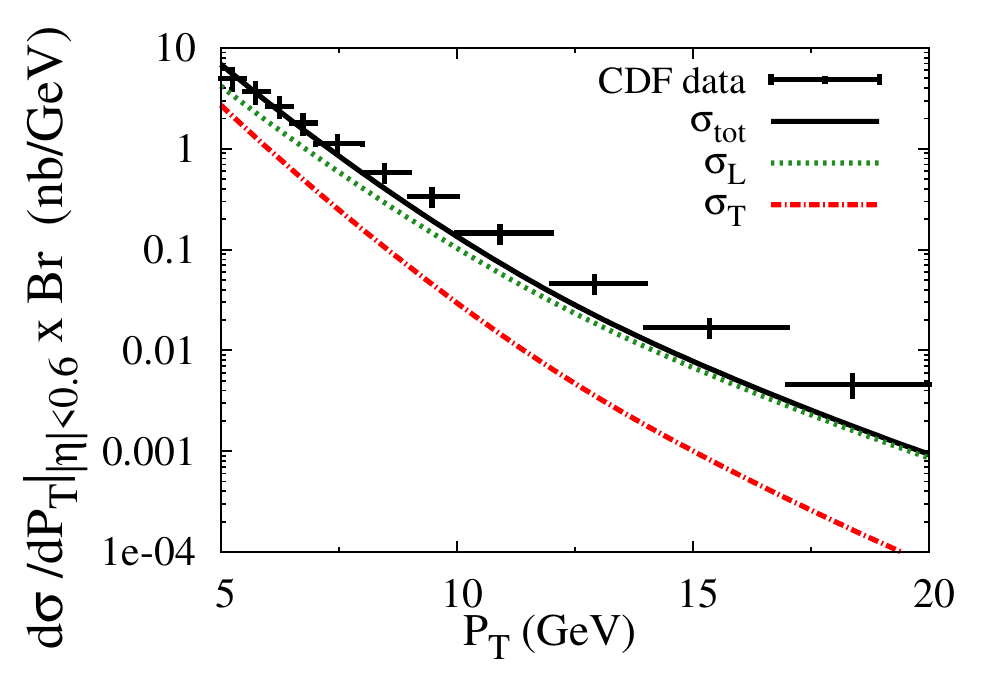}
\caption{Comparison between polarised ($\sigma_T$ and $\sigma_L$) and
unpolarised ($\sigma_{{\rm tot}}$) cross sections [with parameters $a=4$,
$\kappa=4.5$ GeV in Eq.~(\ref{eq:interpolate})], and CDF
experimental data~\cite{Abe:1997yz} at the Tevatron ($\sqrt{s}=1.8$ TeV,
pseudorapidity $|\eta|<0.6$).} \label{fig:cross-sections_CDF}
\end{figure}

\begin{figure}[t!]\centering
\includegraphics[width=.47\textwidth,angle=90]{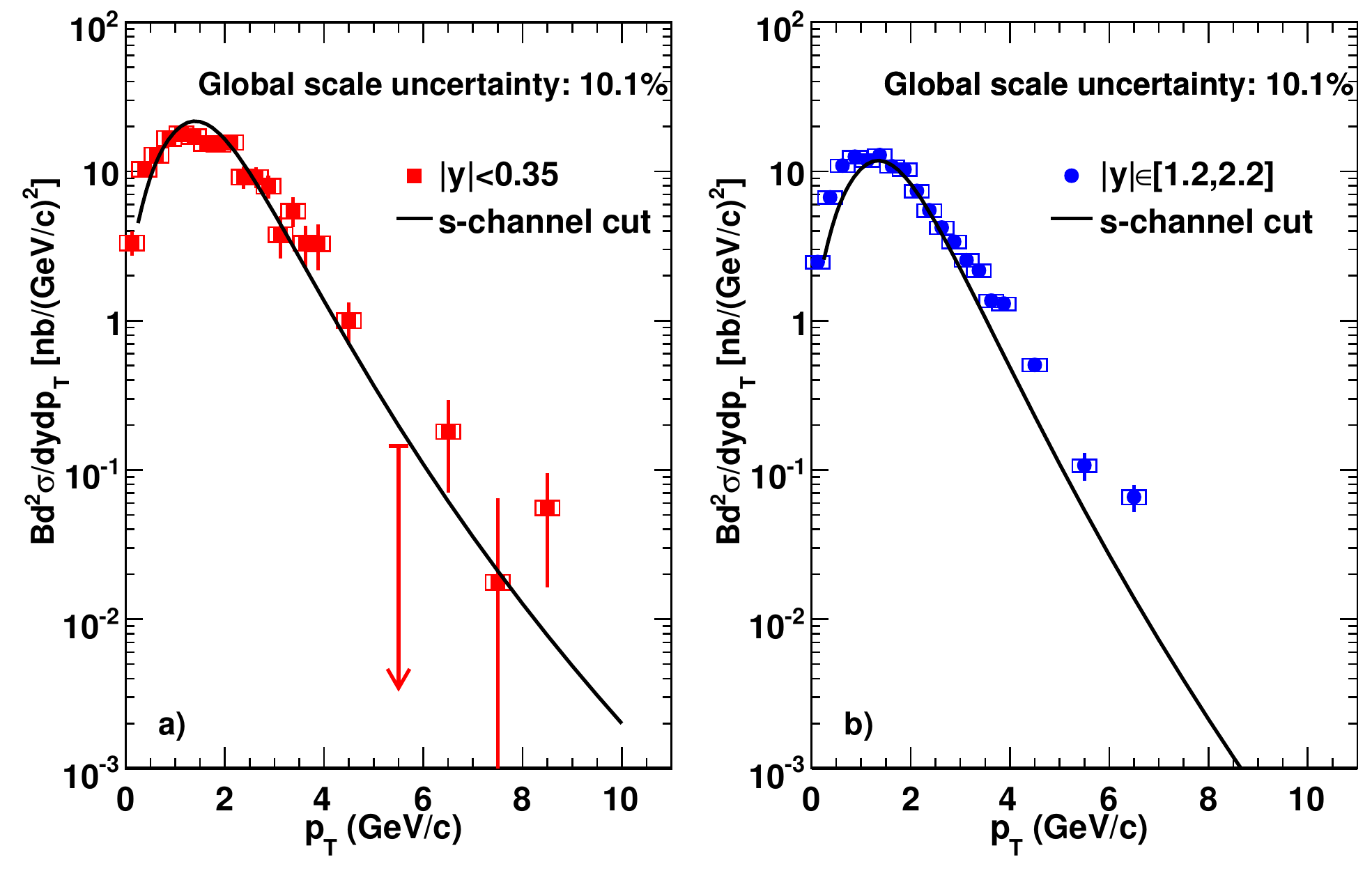}
\caption{Comparison between $\sigma_{{\rm tot}}$ and PHENIX data~\cite{Adare:2006kf} at RHIC
($\sqrt{s}=200$ GeV).
(a) in the rapidity range $|y|<0.35$, (b) $1.2<|y|<2.2$.} \label{fig:cross-sections_RHIC}
\end{figure}

Figure~\ref{fig:cross-sections_CDF} shows our results with parameter values
$a=4$ and $\kappa=4.5$\,GeV for $\sqrt{s}=1.8$ TeV in the pseudorapidity range
$|\eta|<0.6$. The values of $a$ and $\kappa$ were chosen to reproduce the cross-section measurement of direct $J/\psi$ by
CDF~\cite{Abe:1997yz} up to about $P_T=10$ GeV. At higher $P_T$, our curve falls below the data
as expected from the genuine ${1}/{P^8_T}$ scaling of a LO box diagram.
We expect higher-order corrections incorporating fragmentation-type
topologies ($\sim{1}/{P^4_T}$)~\cite{Artoisenet:2008fc,pierre-HLPW08} and associated-production channels
to fill the gap between data and theory at high $P_T$~\cite{Artoisenet:2007xi}.
Figures~\ref{fig:cross-sections_RHIC} show
our results at $\sqrt{s}=200$\,GeV, still with  $a=4$ and $\kappa=4.5$\,GeV,
compared with the PHENIX data\footnote{Note that the PHENIX analysis deals with the {\it total} $J/\psi$ yield, whereas our computation
is for the {\it direct} yield. For the PHENIX kinematics, the $B$ feeddown can be safely neglected. To what concerns the feeddown
from $\chi_c$, it is likely to affect the polarisation observable $\alpha$ (see later), but normally much less the $P_T$ dependence.}~\cite{Adare:2006kf} for the central rapidity region
$|y|<0.35$ (a) and the forward one $1.2<|y|<2.2$ (b).

\begin{figure}[h!]\centering
\includegraphics[width=.75\columnwidth,clip=true]{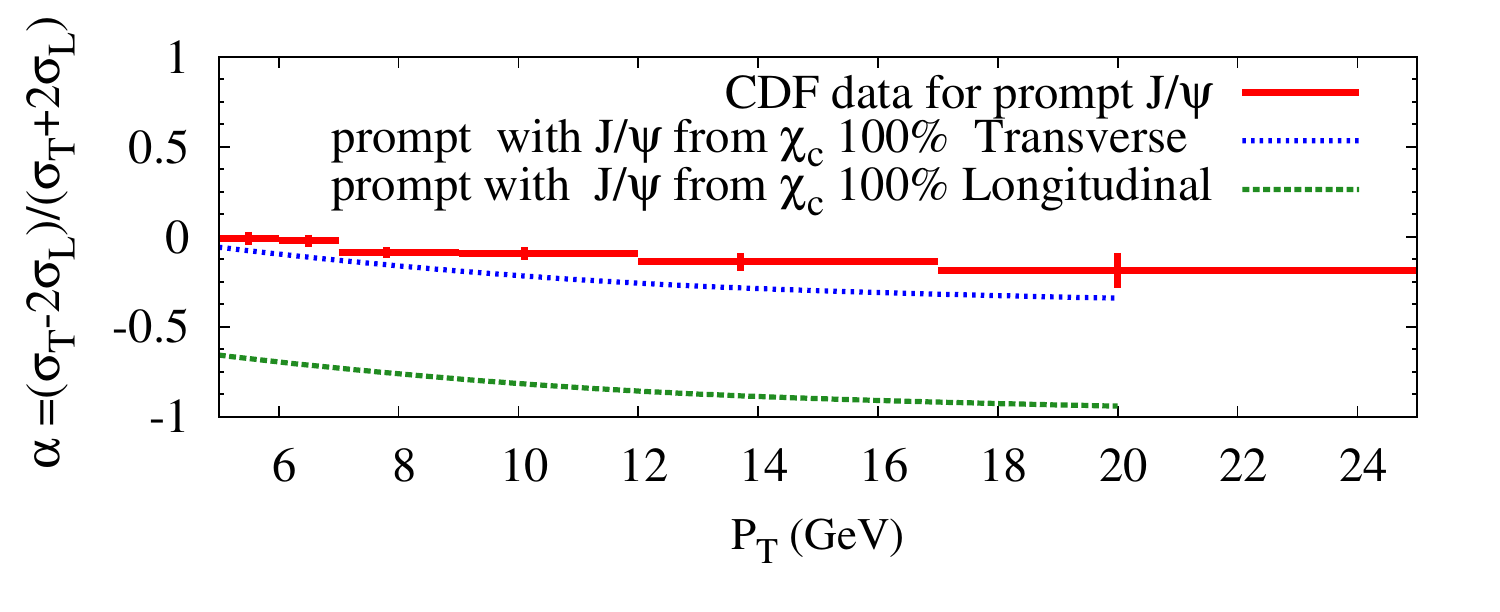}
\caption{Prompt $J/\psi$ polarisation: theory \textit{vs.} CDF
data~\cite{Abulencia:2007us}.} \label{fig:polarisation_CDF}
\end{figure}

$J/\psi$ polarisation measurements at the Tevatron exist only for the prompt yield, we
have thus computed $\alpha$ from our direct-$J/\psi$ cross sections in two extreme
cases, one where the $J/\psi$'s from $\chi_c$ are 100\% transverse and another
where they are 100\% longitudinal, the first scenario being the more likely
one. Figure~\ref{fig:polarisation_CDF} shows the comparison between this
computation and the recent results by CDF at $\sqrt{s}=$ 1.96
TeV~\cite{Abulencia:2007us}. \cfs{fig:polarisation_RHIC} show the
polarisation of the direct yield for the central and forward rapidity regions at RHIC. At very small $P_T$, 
the $J/\psi$ is found to be rather transversal, $\alpha$ being systematically larger for larger rapidity.

\begin{figure}[h!]\centering
\includegraphics[height=.45\columnwidth,angle=90]{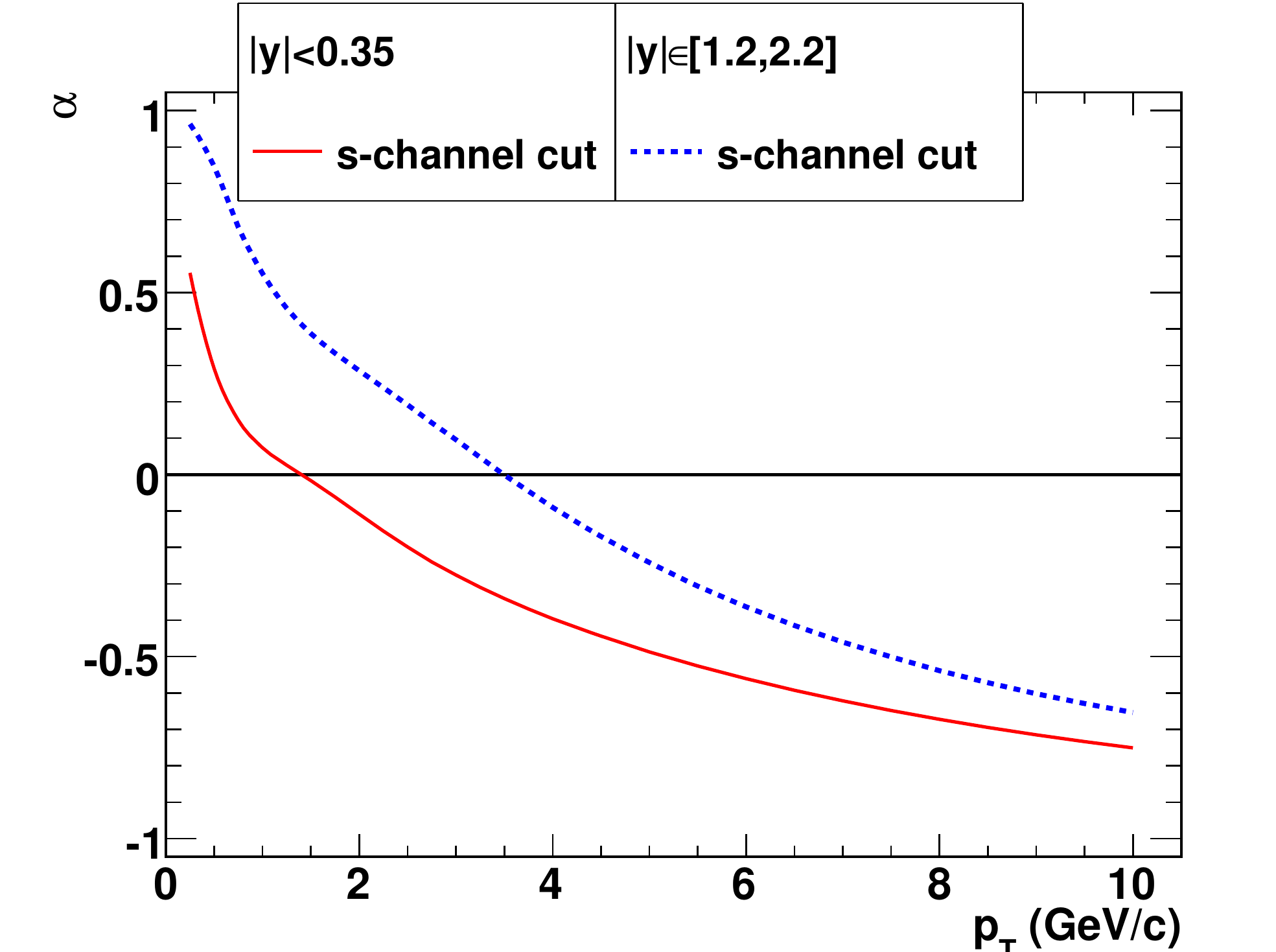}
\includegraphics[height=.45\columnwidth,angle=90]{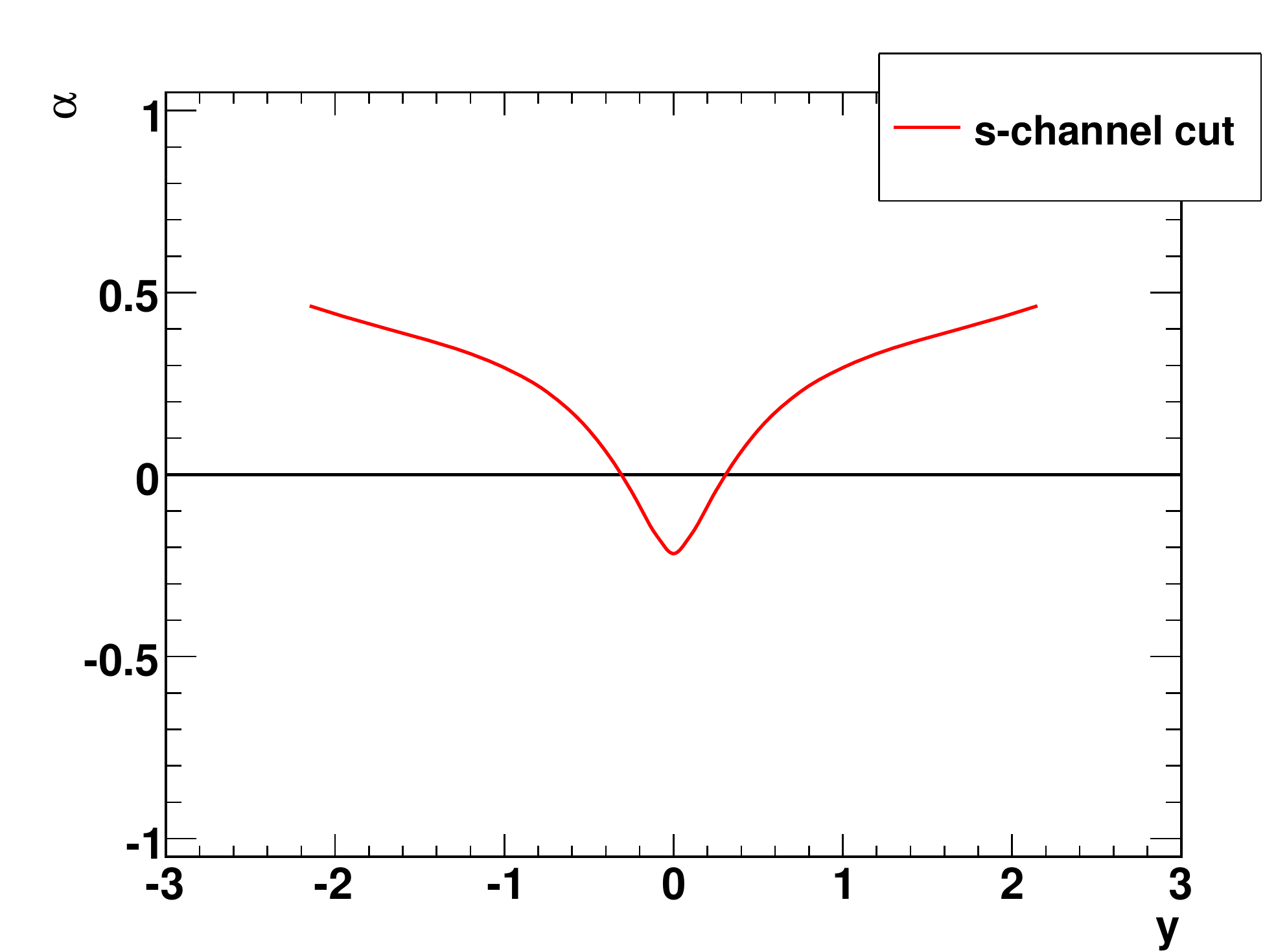}
\caption{(Right) Direct $J/\psi$ polarisation as a function of $P_T$ for the RHIC kinematics in the rapidity intervals
$|y|<0.35$  and $1.2<|y|<2.2$. (Left) Direct $J/\psi$ polarisation integrated over $P_T$  as a function of the rapidity for the RHIC kinematics.} \label{fig:polarisation_RHIC}
\end{figure}

\section{Conclusion}

In~\cite{Lansberg:2005pc}, we showed that there exist two singularities contributing
to the imaginary part of the amplitude for $gg \to J/\psi g$. The first can be identified
to the CSM contribution when the static limit is taken (no relative momentum between quarks).
The second can be referred to as an $s$-channel cut and was never considered 
before~\cite{Lansberg:2005pc}.

To deal with such configurations, we have to introduce a four-point function $c-\bar c-J/\psi-g$,
complementing the information given by the three-point function (or Bethe-Salpeter amplitude)
$c-\bar c-J/\psi$. Such a four-point function is a priori constrained by a low-energy limit 
(when the emitted gluon is soft) and a scaling limit (when the emitted gluon is hard). Given those
two physical constraints, we constructed a four-point functions exhibiting a dependence only on two
parameters, which we fixed to reproduce the Tevatron measurements up to mid $P_T$. 
We then used the latter to compute the cross section for the RHIC kinematics, 
for which we obtained a striking 
agreement with PHENIX data. This agreement can be employed~\cite{extr_vs_intr} to 
investigate on the kinematical effects attributable to the final-state-gluon 
emission in studies of shadowing effects on $J/\psi$ production 
in $pA$, $dA$ and $AA$ collisions, in the spirit of the 
study~\cite{Ferreiro:2008qj,Rakotozafindrabe:2008ms}.
 Our prediction for the polarisation for the prompt  $J/\psi$  yield 
at mid $P_T$ at the Tevatron is mostly longitudinal.

In the COM, colour-octet matrix elements account for
transitions between a coloured heavy-quark pair into a quarkonium by soft unseen
gluon emissions in the final state. In the present approach, the 4-point
function accounts for gluon exchanges between the heavy quarks emitting the
final-state gluon. As for the matrix elements of NRQCD, which are unknown and
then fit, we fixed the unconstrained parameters of this function in order to
reproduce the experimental data at $\sqrt{s}=1.8$ TeV from the CDF
collaboration at the Tevatron for $P_T\lesssim 10$ GeV.

Contrary to usual results obtained with LO calculations, our approach agrees with data down to very low
values of the transverse momentum without need of resummation of initial-state gluon effects. 
This feature could be attributed to the threshold associated with the  cut in the $s$-channel and should
be analysed in more details in the future.

Before drawing further conclusions, several points have to be addressed: Firstly, the size of 
the real part of 
the amplitude has to be evaluated. When fixing the parameter of our four-point function to describe the 
Tevatron data we have implicitly assumed that such a real part was small; this has to verified. Secondly, 
the four-point function we proposed here has to be applied to other regimes of production: a 
similar enhancement by inclusion of the $s$-channel cut is expected in all
production processes where the $J/\psi$ is associated with a gluon,
\textit{e.g.},~photon-photon collision at LEP as well as in photo- and
lepto-production at HERA. On the other hand, other observables insensitive to the COM or the $s$-channel cut 
-- and thus to the ambiguity attached to the description of the four-point function-- should be studied in 
the future, especially at the LHC. To conclude, let us mention two promising new observables, $J/\psi$ production in association with a $c\bar c$ pair~\cite{Artoisenet:2007xi} and the hadronic activity around the 
$J/\psi$~\cite{kraan}.

\section*{Acknowledgments}

J.P.L. thanks J.R. Cudell and Yu.L. Kalinovsky  for our collaboration at the early stage 
of this work, as well as F. Fleuret, S. Peign\'e, H.J. Pirner
 for discussions. Special thanks to  A.~Rakotozafindrabe for providing me with artworks and 
her careful reading of this manuscript.


\begin{thebibliography}{99}

\bibitem{Abe:1997jz}
 F.~Abe \textit{et al.}  [CDF Collaboration],
 Phys.\ Rev.\ Lett.\  \textbf{79},  (1997) 572.
 
\bibitem{Abe:1997yz}
 F.~Abe \textit{et al.}  [CDF Collaboration],
 Phys.\ Rev.\ Lett.\  \textbf{79},  (1997) 578.

\bibitem{review}
 J.\,P.~Lansberg,
 Int.\ J. Mod.\ Phys.\ A \textbf{21},  (2006) 3857;
 N.~Brambilla \textit{et al.},
 CERN 2005-005, hep-ph/0412158;
 M.~Kramer,
 Prog.\ Part.\ Nucl.\ Phys.\  \textbf{47},  (2001) 141.
  %%CITATION = IMPAE,A21,3857;%%

%\cite{Abulencia:2007us}
\bibitem{Abulencia:2007us}
  A.~Abulencia {\it et al.}  [CDF Collaboration],
  %``Polarisation of $J/\psi$ and $\psi_{2S}$ mesons produced in $p \bar{p}$
  %collisions at $\sqrt{s}$ = 1.96-TeV,''
  Phys.\ Rev.\ Lett.\  {\bf 99} (2007) 132001
%  [arXiv:0704.0638 [hep-ex]].
  %%CITATION = PRLTA,99,132001;%%


\bibitem{Affolder:2000nn}
 A.\,A.~Affolder \textit{et al.}  [CDF Collaboration],
 Phys.\ Rev.\ Lett.\  \textbf{85},  (2000) 2886.

\bibitem{Bodwin:1994jh}
  G.\,T.~Bodwin, E.~Braaten, and G.\,P.~Lepage,
  Phys.\ Rev.\ D \textbf{51},  (1995) 1125
  [Erratum, \textit{ibid.} \textbf{55},  (1997) 5853].

\bibitem{Campbell:2007ws}
 J.~Campbell, F.~Maltoni, and F.~Tramontano,
 Phys.\ Rev.\ Lett.\  \textbf{98},  (2007) 252002.

\bibitem{Nayak1}
 G.~C.~Nayak, J.~W.~Qiu, and G.~Sterman,
 Phys.\ Rev.\ D \textbf{74},  (2006) 074007;
 \textit{ibid.} \textbf{72},  (2005) 114012; Phys.\ Lett.\ B \textbf{613}, 
 (2005) 45.



%\cite{Nayak:2007mb}
\bibitem{Nayak:2007mb}
  G.~C.~Nayak, J.~W.~Qiu and G.~Sterman,
  %``Color Transfer in Associated Heavy-Quarkonium Production,''
  Phys.\ Rev.\ Lett.\  {\bf 99} (2007) 212001;
%  [arXiv:0707.2973 [hep-ph]].
  %%CITATION = PRLTA,99,212001;%%
%\cite{Nayak:2007zb}
%\bibitem{Nayak:2007zb}
%  G.~C.~Nayak, J.~W.~Qiu and G.~Sterman,
 \textit{ibid.},
  %``Color Transfer Enhancement for Heavy Quarkonium Production,''
  Phys.\ Rev.\  D {\bf 77} (2008) 034022
%  [arXiv:0711.3476 [hep-ph]].
  %%CITATION = PHRVA,D77,034022;%%

%\cite{Artoisenet:2008fc}
\bibitem{Artoisenet:2008fc}
  P.~Artoisenet, J.~Campbell, J.~P.~Lansberg, F.~Maltoni and F.~Tramontano,
  %``Upsilon production at the Tevatron and the LHC,''
  arXiv:0806.3282 [hep-ph].
  %%CITATION = ARXIV:0806.3282;%%


%\cite{Acosta:2001gv}
\bibitem{Acosta:2001gv}
  D.~E.~Acosta {\it et al.}  [CDF Collaboration],
  %``Upsilon production and polarization in $p\bar{p}$ collisions at
  %$\sqrt{s}=$ 1.8-TeV,''
  Phys.\ Rev.\ Lett.\  {\bf 88} (2002) 161802.
  %%CITATION = PRLTA,88,161802;%%



 %\cite{Abazov:2005yc}
\bibitem{Abazov:2005yc}
  V.~M.~Abazov {\it et al.}  [D0 Collaboration],
  %``Measurement of inclusive differential cross sections for Upsilon(1S)
  %production in p anti-p collisions at s**(1/2) = 1.96-TeV,''
  Phys.\ Rev.\ Lett.\  {\bf 94} (2005) 232001
  [Erratum-ibid.\  {\bf 100} (2008) 049902].
%  [arXiv:hep-ex/0502030].
  %%CITATION = PRLTA,94,232001;%%

\bibitem{CSM_hadron}
 C-H. Chang,
 Nucl.\ Phys.\  B \textbf{172},  (1980) 425;
 R. Baier and R. R\"uckl,
 Phys.\ Lett.\ B \textbf{102},   (1981) 364; Z. Phys.\ C \textbf{19},  (1983) 251.



%\cite{Lansberg:2005pc}
\bibitem{Lansberg:2005pc}
  J.~P.~Lansberg, J.~R.~Cudell and Yu.~L.~Kalinovsky,
  %``New contributions to heavy quarkonium production,''
  Phys.\ Lett.\  B {\bf 633} (2006) 301.
%  [arXiv:hep-ph/0507060].
  %%CITATION = PHLTA,B633,301;%%



\bibitem{Drell:1971vx}
 S.\,D.~Drell and T.\,D.~Lee,
 Phys.\ Rev.\ D \textbf{5},  (1972) 1738.

\bibitem{Kazes:1959}
 E.~Kazes,
 Nuovo Cimento \textbf{13},  (1959) 1226.

\bibitem{Haberzettl:1997jg}
 H.~Haberzettl,
 Phys.\ Rev.\  C \textbf{56},  (1997) 2041.

\bibitem{Haberzettl:1998eq}
 H.~Haberzettl, C.~Bennhold, T.~Mart, and T.~Feuster,
 Phys.\ Rev.\  C \textbf{58},  (1998) R40.

\bibitem{Davidson:2001rk}
 R.\,M.~Davidson and R.~Workman,
 Phys.\ Rev.\  C \textbf{63},  (2001) 025210.

\bibitem{Haberzettl:2006fsi}
 H.~Haberzettl, K.~Nakayama, and S.~Krewald,
 Phys.\ Rev.\ C \textbf{74},  (2006) 045202.




%\cite{Haberzettl:2007kj}
\bibitem{Haberzettl:2007kj}
  H.~Haberzettl and J.~P.~Lansberg,
  %``Possible solution of the J/psi production puzzle,''
  Phys.\ Rev.\ Lett.\  {\bf 100} (2008) 032006
%  [arXiv:0709.3471 [hep-ph]].
  %%CITATION = PRLTA,100,032006;%%

\bibitem{Ohta:1989ji}
 K.~Ohta,
 Phys.\ Rev.\  C \textbf{40},  (1989) 1335.



\bibitem{Lambda}
 M.~A.~Ivanov, J.~G.~Korner and P.~Santorelli,
  %``Semileptonic decays of B/c mesons into charmonium states in a relativistic
  %quark model,''
  Phys.\ Rev.\ D {\bf 71} (2005) 094006;
%  [arXiv:hep-ph/0501051],
  %%CITATION = HEP-PH 0501051;%%
  Phys.\ Rev.\ D {\bf 70} (2004) 014005;
%  [arXiv:hep-ph/0311300],
  %%CITATION = HEP-PH 0311300;%%
  Phys.\ Rev.\ D {\bf 63} (2001) 074010;
%  [arXiv:hep-ph/0007169];
  %%CITATION = HEP-PH 0007169;%%
 M.~A.~Nobes and R.~M.~Woloshyn,
  %``Decays of the B/c meson in a relativistic quark-meson model,''
  J.\ Phys.\ G {\bf 26} (2000) 1079.
%  [arXiv:hep-ph/0005056].
  %%CITATION = HEP-PH 0005056;%%




%\cite{Pumplin:2002vw}
\bibitem{Pumplin:2002vw}
  J.~Pumplin, D.~R.~Stump, J.~Huston, H.~L.~Lai, P.~Nadolsky and W.~K.~Tung,
  %``New generation of parton distributions with uncertainties from global  QCD
  %analysis,''
  JHEP {\bf 0207} (2002) 012.
%  [arXiv:hep-ph/0201195].
  %%CITATION = JHEPA,0207,012;%%


\bibitem{pierre-HLPW08}
  P.~Artoisenet, {\it QCD corrections to Heavy Quarkonium
Production}, this volume.
 

%\cite{Artoisenet:2007xi}
\bibitem{Artoisenet:2007xi}
  P.~Artoisenet, J.~P.~Lansberg and F.~Maltoni,
  %``Hadroproduction of J/psi and Upsilon in association with a heavy-quark
  %pair,''
  Phys.\ Lett.\  B {\bf 653} (2007) 60.
%  [arXiv:hep-ph/0703129].
  %%CITATION = PHLTA,B653,60;%%

\bibitem{Adare:2006kf}
 A.~Adare \textit{et al.}  [PHENIX Collaboration],
 Phys.\ Rev.\ Lett.\  \textbf{98}, (2007) 232002.


\bibitem{extr_vs_intr}
  E.~G.~Ferreiro, F.~Fleuret, J.P.~Lansberg and  A.~Rakotozafindrabe, in progress.

%\cite{Ferreiro:2008qj}
\bibitem{Ferreiro:2008qj}
  E.~G.~Ferreiro, F.~Fleuret and A.~Rakotozafindrabe,
  %``Transverse momentum dependence of J/psi shadowing effects,''
  arXiv:0801.4949 [hep-ph].
  %%CITATION = ARXIV:0801.4949;%%

%\cite{Rakotozafindrabe:2008ms}
\bibitem{Rakotozafindrabe:2008ms}
  A.~M.~Rakotozafindrabe,
  %``A look at the influence of the $J/\psi$ transverse momentum on shadowing,''
  arXiv:0806.3678 [hep-ph], this volume.
  %%CITATION = ARXIV:0806.3678;%%

\bibitem{kraan}
A.C. Kraan, {\it Experimental Aspects of Heavy
Quarkonium Production at the LHC}, this volume.

\end{thebibliography}
\end{document}